\documentclass{article}
\begin{document}

UNIFIED  GAUGE  FIELD  THEORY  AND  TOPOLOGICAL  TRANSITIONS

AJAY  PATWARDHAN

Physics Department ,St Xavier's College,

Mahapalika Marg, Mumbai, India

ajay@imsc.res.in 

Visitor, Institute of Mathematical sciences,

Chennai, India

ABSTRACT

The search for a Unified description of all interactions has created many developments of mathematics and physics. The role of geometric effects in the quantum theory of particles and fields and spacetime has been an active topic of research. This paper attempts to obtain the conditions for a Unified Gauge field theory , including gravity.

 In the Yang Mills type of theories with compactification from a 10 or 11 dimensional space to spacetime of 4 dimensions , the Kaluza Klein and the Holonomy approach has been used. In the compactification of Calabi Yau spaces and sub manifolds the Euler number Topological index is used to label the allowed states and the transitions. With a SU(2) or SL(2,C) connection for gravity and the U(1)*SU(2)*SU(3) or SU(5) connection for the other interactions, a Unified gauge field theory is expressed in the 10 or 11 dimensional space.

 Partition functions for the sum over all possible configurations of sub spaces labeled by Euler number index and Action for gauge and matter fields are constructed. Topological Euler number changing transitions that can occur in the gauge fields and compactified spaces and their significance are discussed. The possible limits and effects of the physical validity of such a theory are discussed.

I  \quad INTRODUCTION

 The geometrisation of physics  program of H. Weyl for gauge groups and A. Einstein for fundamental interactions and spacetime preceeded many of the developments of 'unified field theory'. Subsequently the search for fundamental interactions and theory and experiments have become a still to be completed program. The Unification in terms of Yang Mills gauge fields  described by M.F. Atiyah and E. Witten  in $ 10 or 11$ dimensional space has a grand unified theory ( GUT) with $ SU(5)$ gauge group and $ 4$ dimensional spacetime with $ SL(2,C)$ or $SO(3,1)$ group. The GUT scale and Planck scale are seen to be  of the  same order on a log scale , and all the quark and lepton families and the gauge bosons and the Higgs particle along with the supersymmetric partners are included. To include gravity in the same scheme of Yang Mills gauge fields , the $ SL(2.C)$ is taken as a gauge group and its related group provides a $ SU(2)$ connection, but this approach is not completed yet.

The compactifications over the gauge groups acting on the 10 dimensional space are the basis of a description of fundamental interactions in spacetime. In recent times Path integrals and  partition functions inclusive of topological /geometric indices can be written for condensed matter systems, field theoretic and gravitational systems.

 Index invariant and index changing transformations can arise due to change of dimensions and of the topology /geometry of the space on which the dynamical variables are defined. The dynamics and symmetries of the system with the action and field equations are compatible with the geometry of the manifolds or other spaces. In string theory/ Brane theory/M theory  and gauge field theory , the consistency of Calabi-Yau and other compactification spaces ,  and  of the dynamics on them for the physically relevant objects is required. The compactifications can be written in the Kaluza Klein way of inserting  gauge potentials in the metric of the n dimensional space, or it can be expressed as a Holonomy of  the gauge fields on loops in the subspaces.

 A $ 3+1+6 $ dimension space and the allowed  compctifications to $ 3+1$ spacetime can consistently give the required limits for high energy physics. In condensed matter the local variations in geometry/symmetry and topology are studied for their role in existence of many phases and their  phase transitions. In Yang Mills gauge field theories the unification program in terms of Holonomy operators and the partition functions over the  configurations gives the possibilities of high energy interactions in which there are changes in  topology, that is changes in the Indexes and invariants of equivalence classes of the fields , thus topological phase transitions.

 In the classical and quantum gravity of horizons of black holes and in the pre inflationary cosmological models , the possibility exists of topological phase transitions giving rise to thermal effects , and to  dark matter and energy.There is also a convention of taking a 11 dimensional theory with the $ 3+1+1+2+4 =11$ split in the spacetime and gauge dimensions. There is a neccessity to recover a time dimension for doing the dynamics of entities in the space . Hence the signature of the space has to have a locally Lorentzian form. In 11 dimensions it could be described locally by the representations of $ SO(10,1)$. $ SU(5)$ is a subset of $ SO(10)$; and $SO(3,1)$ is also a subgroup of $SO(10,1)$.

II \quad THE   MODEL  FOR  THE  UNIFIED  GAUGE  FIELD  THEORY AND COMPACTIFICATIONS

The model for the unified gauge field theory is chosen keeping the developments in quantum gravity, M theory and Yang Mills field  theories of the past decades in view. A 10 dimensional space ,  with four for spacetime and gravitation and  six for  the electromagnetic,weak and strong interactions . All these interacting fields are taken as Yang Mills gauge fields with the gauge groups $  SU(2) $ as a subgroup of $ SL(2,C) $ which gives  the local Lorentz connection for gravitation and the $U(1)*SU(2)*SU(3)$ as a subgroup of$ SU(5)$  Yang Mills  gauge theory for the other three interactions in the standard model of particle physics.

In the contemporary way of writing field theories globally with a holonomy loop representation the counting of loop dimensions is done in the following way. Any representative element of the groups if taken in diagonal form would give one for $ U(1)$ , two for$ SU(2)$, and three for$ SU(3)$ as the independent entries on the diagonal.; seen as rank of matrix or number of eigenvalues or elements in the trace .The number of elements would be $ 1+3+8$ for the  individual groups of the product group, but the underlying space is of dimension $ 1+2+4 =7$. Ten or eleven dimensional spaces are frequently used as a standard model coming down from string theory to gauge fields. With $ dim SO(11)/SO(10) = 10$ , a 10 dimensional space for the theory has a  space for gauge compactifications is six dimensions, and the four dimensional spacetime manifold is also carrying a gravity gauge group for unification.

 The light cone and null hypersurface has future and past halves which has a$ SU(2$) group each. Notice that the point is written as $ x(+,-)iy $  and $ z(+,-)ct$ in spacetime as a $ 2*2$ matrix on which $ SL(2,C)$ acts. The point can now be expressed as a Pauli matrix spinor times the  position 4 vector. That is a general elemnt for $ SU(2)$. In the time reversed combination $ z(-,+)ct$ the other option of $ SU(2)$ is realised. This interpretation may avoid the difficulties raised about whether the connection is real or complex for gravitation. Counting two each as dimensions for the two $ SU(2)$ , the 4 dimensions for spacetime and gravitation are obtained . Lorentz  connection based gauge  transformed gravity is diffeomormphed spacetime. Thus the classical idea of deformed or curved spacetime as a representation of the gravitational effect is restored. This could be a  possible argument for the choice.

 In the Kaluza Klein Holonomy approach ; each dimension is represented by a loop; and thus a total of ten loops for ten dimensions. The weakness of gravity creates loops that are macroscopic . If gravity becomes stronger in a unified theory at energies close to the Planck scale then the loops would shrink too and at the beginning of the universe the very small radius of curvarure may go along with the compactified loop getting very small for the gravitational connection. In the late universe we live in the gravitational interaction is very weak and the geometry is classical with a very large radius of curvature and loop size. In the Einstein- Hilbert  action the Ricci scalar is roughly inverse of the square of the radius of curvature; and weak gravity  effects are a nearly flat spacetime, or very large radius of curvature.

 This estimate is expressed as  $ 1/(g_{YM}^2) \int Tr(F \wedge * F) $ for Yang Mills and as $ 1/\kappa \int R \sqrt g $ for the gravitational action. The Ricci scalar is the product of the principal curvatures in the simple case. In both cases the coupling is related to the field strength and the compactification size. For a very weak gravitational constant the radii of curvatures are very large , while for the  other gauge fields with the coupling term of order one, they have very small compactification loops. With this estimate the unified theory permits all gauge fields to act freely , including gravity. With $8\pi G $ as the gravitational constant and a loop radius l , the relation $(8\pi G)^2 L $ as constant gives a scaling for the two , at weak and strong gravity limits.

 The gravitational interaction has a classical limit which is the observable spacetime, while the electromagnetic interaction being $ U(1)$ has a infinite range gauge particle of zero mass and hence also operates in the whole of the spacetime ; even though its holonomy loop is small locally. The $ SU(2)$ connection of gravity has the difficult task of creating a graviton at one end, and the nonperturbative solutions of classical and quantum gravity at the other end of the energy scale  . Spin two representations exist for the $ SL(2,C)$ group and that is still the local gauge group. The $ SU(2)$ subgroup and its connection may need a fresh approach if the Unification has to include the gravitational connection too. Should $ SU(2)*SU(2)$ be considered.

 Consider the group relations : SU(2) $\simeq$ SO(3) ; SU(2)$\otimes$C = SL(2,C) ,

 SU(2)$\otimes$ SU(2)$ \equiv$ SO(3) and   SO(3)$ \subset$ SO(3,1) 

SO(3,1)$ \equiv$ SL(2,C) ; SU(2)= U(2)$\cap$ SL(2,C)

 From these relations the correct nature of the gauge group for gravity has to be determined that encodes all the neccessary information from the Lie group to the Lie algebra , whose representations will give the spinors that go with the connection form. This choice should remain valid for small and large energies and curvatures; that is determine the neighbourhood for the group action on the underlying space.

 The loops for the other interactions  have , due to their larger coupling constants than for gravitation , very small sizes. How does the partition of the ten dimensional space occur so that it is $ 4+6= 10$. What are the allowed compactifications and what is the topology of these loops in terms of Homotopy . What are the allowed differential forms and hence the gauge fields on this space. And hence what is the cohomology of closed and exact forms that gives the Betti numbers and from which the Euler number can be defined. What indexes dependent on the field forms could be defined that are topological invariants ,  that set up a classification of the field equations, compatible with the geometry and topology of the subspaces of lower dimensions of full space of 10 dimensions.

 For a even dimensional manifold with a nowhere vanishing vector field ,that exists as gravity is universally present ; the Euler number is zero. This is true for the 3+1 dimensional spacetime , subspace of the ten dimensional space and for that 10 dimensional space itself. The gravity equations giving dynamics of three surfaces will have Euler number zero for the odd dimensions , unless closed time like loops or singularities are present. The number of iso spectral Riemann surfaces of genus g grows exponentially with the square of g. Hence genus changing transformations of spacetime will have to be exceptionally handled. The Gravitational field equations constrain the spacelike surfaces to evolve in such a manner that the 4 and lower dimensional manifolds will have genus zero or 1. This is not topologically complicated.

III \quad PARTITION  FUNCTIONS  AND TOPOLOGICAL  INDEX

The Partition function methods of statistical mechanics as applied to field theory are used. In the presence of symmetry groups in molecular and solid state physics that give rise to allowed configurations; the partition functions have to sum over all the configurations , with the Boltzman or Gibbs distribution and density of states as the weight function .

 The probability for a configuration and the average values calculated with it give the thermodynamic quantities. Thus for example the partition function   $ \Sigma_J (2J+1) \exp((-J(J+1))/2IkT) $ has the sum over $ J $. The $2J+1 $ possibilities (degeneracy) for the quantum number or index $ J $ are summed . If any selection rules were to create allowed  $ J $ values then the number of configurations to be summed over in the partition function would be determined. If the number of independent configurations to be summed over is given by any other rule, then that factor is included in the partition function as the weightage for the ensemble. 

For example if there were $N_i$ configurations of the $i$ th kind , then the sum

 $Z(T)= \Sigma_i N_i exp(-E_i/T) =\Sigma_i exp(-E_i/T +ln(N_i)) = \Sigma_i exp(-F_i/T) $

 with $F_i =E_i-TS_i $ and $ S_i =ln(N_i)$ for the Free energy and entropy and the probabilities given by

$ P_i =( N_i exp(-E_i/T))/Z(T)$ and $ <A> =\Sigma_i P_i A_i$. The configurations can come from the energy levels as well as the symmetries.

 The canonical ensemble $exp(-\beta H) $ Gibbs density goes over to the $exp(iS)$ in path integral form and with action $S $ , and the euclidean form is $ exp(-S) $. This is understood as a functional integral over the matter and gauge fields.

 $ S [\phi_i , A^a_j]$ and $Z = \int d [\phi_i,A^a_j] exp(-S[\phi_i, A^a_j] )$. The configurations could follow a symmetry group or also a topological classification.

 To introduce topological properties in the partition functions ; topological indices such as the Euler number label the equivalence class of subspaces and field configurations compatible with them. These are identified and summed over in the partition function. Thus the

 $ Z(M) = \Sigma dim(M)^{\chi(M)} \int exp(-S)$ , where the sum is over all the possible Euler number topological configurations and the integration is on all the fields in the action S. A index changing transformation will have an effect on the partition function and other quantities. Writing the action integral,path integral form of partition functions  , in perturbatively solvable interactions, the attempt is to obtain gaussian integrals.

 In the modern theory of manifolds and gauge fields compatible with them, the spectrum of differential operators is related to and controlling the allowed geometries. Some invariant indexes are defined. The gauge fields expressed in differential forms will have their integrals define the indices. The closed and exact forms will give the cohomology and the Betti numbers. This defines the Euler number for the 10 dimensional space and the subspaces.

 Consider the set $X_p(M_n) $ of closed forms of order $p$ on the sub manifold $M_n$ of dimension $n$, and the set $Y_p(M_n)$ of exact forms of order $p$. Then the quotient space $X_p/Y_p $ gives the cohomology $ H_p(M_n)$. The dimension of this is the Betti number . $ dim (H_p(M_n)) = b_p(M_n)$. Consider the collection of all possible $p$ order differential forms ; in this case the gauge fields and potentials and related forms. Then the Euler number characteristic index is defined by the sum $ \chi(M_n) =\Sigma_{p=1}^n (-1)^p b_p(M_n) $ for the subspace $M_n$ of dimension $n$. Consider the collection of subspaces of all allowed dimensions and the equivalence class of all such subspaces having the same Euler number and dimension ; this describes the ``degenerate excitation spectrum'' of the theory.

 It is connected to the gauge field theory chosen and the underlying geometry and topology of the subspaces or moduli spaces of the 10 or 11 dimensional space .In the simple case of 2 dimensional Riemann spaces the Euler index is given by the formula $2-2g$ where $g$ is the genus, typically the number of holes minus the number of handles minus the number of boundaries. A simple model of a  gaussian integral for action of the matter and gauge fields on the Riemann space with $n$ holes gives a partition function that can be summed as a Series ; in terms of Riemann zeta functions.

IV  \quad TOPOLOGICAL  PROPERTIES  OF  THE  SUBMANIFOLDS AND TOPOLOGICAL TRANSITIONS

 The fundamental interactions of GUT form  6 dimensional  subspaces or loop configurations that can have non trivial topologies. The Calabi-Yau spaces also provide a reduction of the 10 dimensional space to the lower dimension ones with various  Euler numbers. As the genus number increases the number of possible configurations with the same genus number (an equivalence class) increases rapidly. Compatibility conditions of the fields and the dynamics will reduce the number of allowed configurations. In the partition function, the sum over all the possible (in)equivalent configurations labelled by the allowed Euler numbers is taken. The 10 dimensional apace has subspaces which could be manifolds of lower dimensions, Calabi -Yau spaces , and the holonomy loops have homotopy groups associated with them.

 The Chern number, topological charge, Atiyah -Singer and Hopf -Poincare and other indexes are used to characterise the topological invariants of the field theory. But this gives the model  and dynamics dependent part of the interactions and it depends on the field configurations compatible with the underlying spaces. The compactification to sub spaces of lower dimensions will lead to equivalent and inequivalent classes under the Euler number index ; and this  is the chosen index in the research literature . It is related to the Betti numbers and hence to the cohomology of the differential forms of fields and the cycles for integration on subspaces. It is also obtained from the Gauss Bonet theorem and the properties of the critical points of vectorfields in terms of the Morse index ; and it is also a well defined number for discrete lattices, graphs and polyhedra. In condensed matter systems the structural phase transitions and discrete lattice topology changes involve the Euler number with number of vertices minus number of edges plus number of faces.

In the picture suggested in this paper, high energy interactions could allow possible changes of topology for the intermediate states of excitations. Symmetry group changes are well known and broken symmetry gives  a number of well tested effects in physics. Topology changing transformations are increasingly being found in a variety of physics. Between the Planck scale and the scale of energies of the experimental data , there is a large range that may give rise to a number of phenomena. In the 10 dimensional unified gauge field theory of all interactions, the program of geometrisation of fundamental physics,  identifies the properties of the  underlying space with the fields present on it using gauge  connection and action ; and compactifications compatible with the dynamics are expected. The spectrum of differential operators is directly connected with the geometric properties of the manifold.

 If Yang Mills theory had been recognised as the underlying theory of fundamental interactions while Einstein was working , then he may have attempted to enlarge his 4+1 five dimensional unified theory of electrodynamics and gravitation ;  such that additioanlly 2+3 dimensions would have been added to include the weak and strong interaction potentials in the metric as a ten*ten matrix. Perhaps realising that the modern theory of differential geometry , having evolved from Riemannian geometry shifts the emphasis from the metric to the connection; a follower of Einstein may then have tried to write both gravitation and the Yang Mills theories in the language of differential forms and spinors when quantum theory too had to be done. In that way the developments could have led to a unified gauge field theory.

 However  the developments  took decades following Yang Mills work to  actually  lead to  Abhay Ashtekar writing the SU(2) and SL(2,C) connection or gauge potential in the spinor form as new variables for gravity. This opened the possibility that all interactions are gauge field theories and the equations for fields would follow from the gauge connection. The Holonomy representation and path integral methods have now created a new possibility for a unified approach and possibly a real unification of all the interactions including gravity.  The SU(2) subgroup of SL(2,C) and the U(1)*SU(2)*SU(3) as a subgroup of SU(5) provide the gauge groups for the four fundamental interactions.

The ten ( or eleven) dimensional space in which the holonomy loop integrals , with these gauge groups , are defind are the theory's basic assumption. The compactification to obtain the reduced four dimensional subspace as the spacetime with a gravitational connection form is the program of those in classical and quantum gravity. The quantum Riemann geometry and loop quantum gravity  refers to this four dimensional  subspace and the macroscopic world arises as a spacetime manifold in the classical case. The reduction of the 6 dimensional subspace of the ten dimensional space, into the Calabi-Yau spaces and manifolds is the subject of high energy physics done with string/brane/M theory or with the Yang Mills gauge field theory in Holonomy loop representation. Non trivial topological properties can arise in the compactifications.

To illustrate the topological trsnsitions that are  possible the path integral over the action is augumented by the sum over configurations labelled by the Euler number index to obtain the partition function. In this an analogy with partition function applications in statistical mechanics of condensed matter systems that give topological phase transitions can be made.The Yang Mills theory in a two dimensional Riemann space with genus g has a partition function  given by  $\Sigma ( dim(M)^{\chi(M)} $ *action integral. The Euler number is 2-2g, here with genus number g. If the Riemann space is a 2 sphere or 2 torus or a collection of torii then the genus number is a easy examole to obtain.

 The general case of any subspace of dimension 6 or less is to be obtained in a similar manner , however the equivalence class of identical Euler number  has a large number of configurations possible. The weightage  for any configuration will be given by the evaluation of the path integral for the action of the  theoretical model of  the interactions . In the simple case when path integral can be written as  gaussian integrals is taken then this gives a dimenesion dependent term and restricts the configurations to those of high symmetry for the field configurations over which the functional integral is defined. The Euler number could be counted or computed in a variety of ways , but taking  the sum of Betti numbers is the preferred choice .  Gauge fields give a natural condition of closed and exact forms and the cohomology for the  upto six dimensional sub spaces of the ten dimensional space. This would be seven dimensional if the 11  dimensional theory is used.

 The Euler index of the sub manifolds $M_n$ of dimension $n$ are also given by the Hopf theorem as : For closed $M_n$ and vector field $v$ on $M_n$ with isolated and finite number $p$ of singularities, the sum $\Sigma J_v(p) =\chi(M_n)$ ; where the $ J_v(p) $ is the index of the vector field $v$ at the  singular point$p$. The singularities of the gauge vector field on $ M_n$ and their indices at these points hence control the euler number. This is an equivalence class as several different choices of $ M_n$ could have the same Euler index. Hence it gives a sum in the partition function over these configurations. The change in the Euler index caused by the various things it depends on, give the ``topologically inequivalent'' spaces. 

 Topological transitions between inequivalent spaces in this sense can occur. The variation of the Euler index as the compactification of Calabi Yau spaces, the interaction model and parameters, and the energy scale are varied could be seen. If the second derivative of the Euler index is calculated then the first and second order topological transitions can be described. In condensed matter physics , a model potential with its critical points varying as a parameter is varied, is used to find the condition $\partial^2 \chi/\partial v^2 > 0 , < 0 $ for the two transitions.

 If this analogy is extended to the Euler number defined with the singularities of the gauge vector fields , then there is a classification of types of topological phase transitions. Consider  the second partial derivative of the Euler index with respect to any pairs of the  gauge fields and the determinant greater than or less than 0. The number of negative eigen values of the Hessian give the index.

V \quad THE EQUATIONS FOR THE UNIFIED GAUGE  FIELD THEORY PARTITION FUNCTIONS WITH EULER NUMBER TOPOLOGICAL  INDEX

The partition function $ Z(M) =\int d[\phi]Exp[-S(\phi)]   $

 is modified to  $\Sigma dim(M)^{\chi(M)}\int d[\phi_i,A_i]exp[-S(\phi_i,A^a_i)] $

 and the Holonomy is given as  $ TrPExp(-ie\oint_ {M(c_i)} dx A^at_a ) ) = W[M_n,A^a_i,C_i]$

 the Holonomy for the full gauge group can be written as product of the Holonomies of the sub gauge groups and the moduli spaces taken for the integrals

$W[M,{A}] = \prod_i W[M_i,A^a_i]$ The product is over all possibble sub manifolds and the gauge connections on them

 This can be also expressed in terms of the expectation value of the field operators in usual field theory $<A(x)---A(x')>_n$

The Euler number is $ \chi(M_n) = \Sigma_{k=0}^n (-1)^k b_k(M_n)$

 with $\chi(M_n\times M_l) = \chi(M_n) \chi(M_l)$

 and the Betti number is  given by the  $dim(H_n) = b_n $

 for the cohomology $H_n$ , the quotient of the set of closed and exact forms on the subspace of order $n $ is needed 

 action $S(\phi_i, A^a_i)$ is for the gauge fields and matter fields dependent on the model of the interactions 

and the dimension of the space has index $i$ from 1 to 10, and the subspaces with the index $n$ from 4 to 9.

For the Yang Mills gauge fields with gauge group represented by the $t_a$ and the connections or gauge potentials by $A^a_i$ 

the fields are $F= dA + A\wedge A $ and the action is $-1/e^2\int d^nx Tr_n(F\wedge *F)$

The standard model or its unificationin SU(5) has the usual Yang Mills connections $A^a_i t_a $ where the $t_a $ are the group representation and the $A^a_i$ are the gauge potentials with group index $a$ and the index for the underlying space as $i$;,that can go upto 10 or 11 dimensions ; but is restricted to 4 after compactification , as the spacetime index. The gauge fields are $F=dA + A\wedge A$ and the action is $S =-1/g_{YM}^2 \int Tr (F\wedge *F)d^nx$ . There is a topological invariant $ \int d^nx Tr(F\wedge F) $ which can be interpreted like a Gauss Bonet term or 2nd curvature form. There is a topological charge $\int d^nx F$ or $\int d^nx*F$ , depending on whether the self dual or anti self dual case is taken, respectively $*F=iF$ and $*F=-iF$. The gauge group acts on the potentials or connections as $A' = gAg^{-1} + g\partial_\mu g^{-1} $ and $ F' = gFg^{-1}$. 

A similar construction would be expected for the gravity gauge group. But there are some differences. In the Sen, Ashtekar version the $  SU(2)$ real and complex cases were taken respectively for a Hamiltonian form of the theory. In A Magnon's covariant and geometric generalisation to include the $ U(1)$ gauge group unified with the $SL(2,C)$, the role of the $ SL(2,C)$ connection becomes unified with the $ U(1)$ . The Einstein Maxwell theory is obtained with a possibility of a Yang Mills form. The action in the Einstein Hilbert form of $\int d^4x\sqrt g R$ arises as a geometric invariant, that is extremal when the consistency condition is satisfied. Namely the Einstein equation for gravitation, with the Yang Mills form of the $ U(1)$ electromagnetic stress energy tensor as  source, and the equations satisfied by the field ; as well as the Bianchi equations.

The SL(2,C) Lorentz connection in spinor variables  is $ A^{IJ}_\alpha $, which is antisymmetric in $I,J$ and has a dual obtained using the $\epsilon^{IJ}_{KL}$  , with the self dual and antiself dual pair $A^{(+,-)} = i/2(A[-,+]i*A)$ . Then the Field or curvature obtained from the connection is given by

$F^+{IJ}_{\alpha\beta} =\partial_\alpha A^{+IJ}_\beta -\partial_\beta A^{+IJ}_\alpha +[A^+_\alpha , A^+_\beta]^{IJ} $ 

 and this is self dual $*F =iF $

Then the tensor quantities for general relativity are found as $g_{\alpha\beta} = \eta_{IJ} e^I_\alpha e^J_\beta$ 
as the metric in terms of the Lorentz frame and Minkowski metric.

 The Christoffel symbol is $\Gamma^\gamma_{\alpha\beta }= A^J_{\alpha I} e^I_\beta e^\gamma_J $ and

 the Riemann tensor is $R^{+\alpha\delta}_{\beta\gamma}  =  F^{+IJ}_{\beta\gamma} e^{\alpha}_I e^{\delta}_J$

From this the Einstein equation is $ R^+_{\alpha\beta} -1/2 g_{\alpha\beta} R^+ =0$

The problem of having a complex connection and non compact gauge group has been discussed but not resolved. There is also the question of having a a Lorentz gauge connection defined like all the other gauge connections on the 10 or 11 dimensional theory; and reducible to the 4 dimensional one for obtaining the classical spacetime.

 This construction in A Magnon's work is  expected to be generalisable to include the other non Abelian gauge field theories too ; but the success of this framework is not established convincingly in the 4 dimensional spacetime. However the possibility of starting with the higher dimensional space for a Unified field theory and then carrying out compactifications in the Holonomy representation for the Gauge fields could yield an alternative method . The Poincare group including the translations is also taken as the gauge group , but that program leads to additional terms in the action. This could be creating an interpretation problem when the combined gauge fields are taken. In this paper it is assumed that the correct gravity gauge group will eventually get fixed among these candidates and the higher dimensional unification will work as a program to get the classical limit as well as the Yang Mills like quantisation of the theory. The basic quantity to calculate is the partition function.

For the compactified subspaces the partition function becomes 

 $ Z(M_i) = \Sigma_i dim (M_i)^{\chi(M_i)} \int_{M_i} d[\phi,A] Exp( -S[\phi,A])$

The compactifications could be expressed in the Kaluza Klein way as  a line element, to illustrate the concept,

$  ds^2 = f(r,t)dt^2 -g(r,t)dr^2 -\Sigma_{\mu,\nu}\Sigma^{dimG(i)} (A^a_\mu  A^b_\nu t_a t_b)_{ij} \zeta^i_\mu \zeta^i_\nu  $;

 $\mu , \nu = 3 $to $ 10 $ or $ 11$

As an example of an Yang Mills partition function in 2 dimensions 

 $ Z = \int DA_\mu exp(1/g^2 \int_M d\mu Tr(F_{\mu \nu} F^{\mu\nu})) $

For a simple model in two dimensions this has been evaluated

 $ Z = \Sigma_R (dimR)^{2-2G} exp(-g^{-2} A C_2(R)/2) $

 where G is genus, g is Yang Mills coupling, A is area of metric on M, C is second Casimir invariant of Riemann space R

$ \chi(M_i)$ are Euler characteristics of Moduli spaces of a genus g Riemann surface ,

 and the $ \chi = 1/8\pi^2 \int d^4x \sqrt g F\wedge F$ is a  Gauss Bonet integrand 

 It can be $\chi = 2-2g-n $ for genus g and n punctures

The compactification is done on the moduli spaces of the 10 or 11 dimensional space due to the gauge sub groups. The Calabi-Yau spaces are the sub manifolds described by the general formula 

 $ \Sigma_{j=1}^n C_{1, j,n} \prod_j \zeta_1^{s_1} \zeta_j^{s_j} \zeta_n^{s_n} = 0$,

 which is a polynomial in products of the variables $\zeta$, or a multidimensional polynomial. Computational packages such as Mathematica can give visualisations of these spaces and a classification of their Euler indices can be used for the equivalence classes.

 When one Calabi-Yau space changes into another in compactification, the exponents and the coefficients in the formula defining it  change. The identical Euler number of a class of such spaces can be used to describe the equivalence class of the spaces. They have dimension upto $ n=10$ and the Euler index . The equivalence class of all such spaces with the same Euler number is taken. The collection of all such equivalence classes labeled by their Euler numbers is the topological excitation spectrum of the theory. The dimension changing transformation is expressed as one or more of the exponents going to zero in the product over the Calabi Yau coordinates in the multi polynomial sum. The sum over all such configurations of each dimension allowed by the gauge  field dynamics and the compatibility of the differential operators and the manifolds they act on is a complicated classification problem at the frontier of mathematics. However it is seen that the dynamics of the fields  and the properties of the manifolds they act on both play an important role.

 As a simple example consider that the sub manifold described as a quadratic expression in three variables is a torus that could be changing into a sphere.  A sphere with two holes and a handle attatched is topologically a torus. The genus number is number  of holes minus number of handles,and is 2-1=1 for a 2-torus and zero for a 2 sphere. Hence the Euler number which is 2-2g for a 2surface changes from 0 to 2 when the torus changes to a sphere, by pinching off the handle ,or the torus along the smaller radius is pinched off. This involves the singular points of the vector field on the surface, and could be also seen as a Holonomy loop being made to disappear from the product of Holonomy loops ,that is become $ e^0 =1$ or $ exp(I)$ as an operator.  
 
The partition function is expressed as

 $ Z = \Sigma_{M(n)=1}^{10} dim (M(n)^{\chi(M(n))} \int_{M(n)}d[\phi_i,A^a_j] Exp(-S[\phi_i,A^a_j]) $

and the expectation values of the Holonomies are taken as

 $ <W[(C)]> =i/Z \int_{G} D[A] W[C,A] exp(-S_{YM}(A))$ on loops $C$

the general result is

$<W(C_!)--W(C_n)>=1/Z \int D[A](W(C_1,A) --W(C_n,A)) Exp(-S_{YM}(A)) $

for the n loops Holonomy  and the  $-1$ has been inserted in the exponent as  replacement of the $2\pi i/h$ as coeffecient of the Action 

$ \int(\phi_j^+ \phi_k) exp(-\phi^+ F \phi)d\phi^+d\phi  = (\int exp(-\phi^+ F \phi)d\phi^+d\phi )) 1/F_{jk}$

 gives for quadratic forms, gaussian like integrals which could be used for model calculations

VI \quad DISCUSSION OF THE UNIFIED GAUGE FIELD THEORY , INCLUDING GRAVITY,  AND TOPOLOGY CHANGING TRANSITIONS

The topology changing transitions are understood as follows. A transformation  manifold $ M_n$ to $ M_n' $ such that  in the definition of the Calabi Yau spaces there is a tranformation of any coefficients and powers of the variables $\zeta$. The new space also has a euler index, either the same or different. A change of dimension by compactification too can occur. The equivalence class of same dimension and Euler number can be found.  Then the total partition function is reevaluated.The simple example is to take the action consisting of quadratic forms in matter and gauge field variables. then the matrix gaussian is the exponential of the action integral and this gives the normalisation factor upon doing the functional integral. There will be a fixed factor dependent on dimension and volume of the space $ M_n $ over which the integral is done. Then the sum over all the $ M_n $ could be performed.

 $ Z = \Sigma _{M_n} ( dim (M_n))^ {\chi(M_n)} f(M_n,G_{M_n}) $

A closed compactified odd dimensional sub manifold makes change of topological index  possible. Consider a congugacy class of loops for Holonomy on the submanifolds with the same Euler number. Then the Kaluza Klein compactification is a topology changing transition. The sum is over all configurations of the allowed subspaces of dimension upto 10 and the equivalence classes of subspaces described by the same Euler indexes for each such dimension. The integral is a path integral or functional integral over all the gauge fields $A^a_j$ and matter fields $\phi_i$ in the action and in the measure.

 The example of the 2 dimensional Yang Mills theory gives a result involving the area and the Casimir invariant. In general in the usual way of integrating over the allowed sub manifolds the exponential is expected to be turned into a multi dimensional gaussian functional and then integrated . This will give a quantity that depends on dimension ,for the volume of sub manifold, and on the  couplings and characteristic Casimirs of the gauge fields; in the summation over the possible configurations in the partition function.

The coordinates in the compactified dimensions are angles , and the loop spaces for compactification have very largre radii for the gravity gauge group and very small radii for the other gauge group connections. As the energy increases to the GUT scale , the three interactions give the single SU(5) ccoupling 0.033. Gravity coupling is expected to be nearly constant till GUT scale of energy is reached and it could rise rapidly thereafter to become like the GUT coupling,  in going towards  the Planck scale. It is expected that in this regime the compactification loop for the SU(2) gravity gauge field will become very small, like that for the other interactions. When considered on a log scale for energy as well as for coupling constant, the variation for the three fundamental interactions, electromagnetic, weak and strong , it is a slow variation. For gravitation the behaviour of the gravitational constant could be flat  or constant almost upto the GUT scale of energy $10^{15} $ GeV and then rise till the  Planck scale of $10^{19} $ GeV; thus going from a value of $-38$ on the log scale to that of $-2$ for the gravitational coupling. This would provide a basis for considering a Unified Gauge Field Theory for all interactions and give a possibility of a consistent picture of Physics.

 But these conditions are likely to occur only near a singularity or in the beginning of the universe. The unified field theory in this regime will have a gauge group that includes the SU(5) and the  SU(2) or SL(2,C) as subgroups. Otherwise the weakness of the gravitational interaction compared to the others by a factor of $ (10^-38)$, allows a compactification  loop almost  $(10^{38}) $ times bigger. This  gives a classical and macroscopic size universe with  General theory of relativity giving its dynamics. On this spacetime manifold the  other gauge fields exist in compactified dimensions and create the physics at all the energy scales below the GUT scale. This is indeed fortunate for all of us living in  the Universe.

The difficulties with the renormalisability of gravity have been given many reasons ; that do not include the possibility of the coupling becoming very large ,like GUT coupling , but only at energy scales beyond GUT scale. If classically gravity can be expressed as a gauge theory how does the breakdown of the perturbative scheme occur in quantised gravity of any form. Does coupling constant renormalisation work. In the non perturbative approaches the emphasis is on loopspaces of holonomies , and these  arise from a string /M theory or from the quantum Riemann geometry. In a path integral form of field theory the additional terms to be put in as perturbative corrections in the action and the gauge fixing terms have been known to give divergent quantities for gravity. How serious is this problem below the GUT scale ; and can it be avoided at the higher scale by the coupling constant increasing rapidly as a  single Grand unified theory couples to gravity.

From the understanding, that fundamental interactions occur in Minkowski spacetime locally, and  the expectation that ,geometric properties of a suitable underlying space should determine the physics globally of all the interactions; which Albert Einstein had in the first half of the 20th century , the program of a unified description of physics has come a long way. The twenty first century has begun with, as yet, no complete theory ; but many of the aspects of the Fundamental theory are known. This paper has discussed some of the possibilities of a Unified Gauge  Field Yang Mills Type of a  theory of all the interactions, and the significance of the topological index transitions for physics.

VII \quad ACKNOWLEDGEMENTS

I thank the Director,Institute of Mathematical Sciences, Chennai ,India and Prof N. D.HariDass  for supporting my visit. I appreciate the  Institute facilities and discussions with its members. My 5 papers were written at the Institute  while on vacation from St Xavier's College, Mumbai, India ; where I have an active  theoretical physics group.

VIII \quad REFERENCES

1.M.F.Atiyah The geometry of Yang Mills fields.Pisa 1979 lectures

2. Collected works of MF Atiyah Gauge theories

3.E.Witten  Unification hep-ph/o207124 in www.arxiv.org

4.E.Witten Quest for unification hep-ph/9812208

5.A.Ashtekar, J Lewandowski Background independent quantum gravity gr-qc/0404018

6.A.Ashtekar Non perturbative quantum general relativity in Ed.Julia and J.Zinn-Justin Les Houches 1992 Gravitation and quantisation.North Holland

7.T.Frankel The Geometry of physics Cambridge university press 1997

8.T.Dass Symmetries,gauge fields,Strings and fundamental interactions Wiley Eastern 1993

9.M.Nakahara Geometry, Topology and Physics Institute of Pjysics 1990

10.M.Blau Quantum Yang Mills theory on arbitrary surfaces 1992

11.O.Alvarez Lectures on quantum Mechanics and the Index theorem.

12.J.Gegenberg and G Kunstatter partition function for topological field theory Ann Phys 1994

13.Aspinwall etal Spacetime topology change In Calabi Yau moduli spaces. hep-th/9311186

14.A.NestorovTopology change in 2+1 dimensional gravity with non abelian Higgs field gr-qc 0403079

15 A. P. Balachandran s.Kurkanoglu Topology change for fuzzy physics hep-th 0310026

16 R.Ionicioiu Topology change from Kaluza Klein dimensions gr-qc/9709057

17. Ed Gerardus t'Hooft 50 years of Yang Mills theory world scientific press 2004

18 A.A.Migdal Loop equations and area laws in turbulence in Quantum field theory and string theory Ed L.Baulien, V.Dtsenko,V.Kazakov,P.Windev Nato ASI seriesv328

19 P . Peldan Phys Rev D 46 , R2279 1992 

20. gr-qc/0402069 , gr-qc 0311036 , hep-th 0401160,hep-ph 0303185 , gr-qc 0204019 , hep-th 0312022 , cond-matt 0105426,cond matt 9904073, cond matt 0309556

21.Formation and interaction of topological defects Ed A.Davis, R. Brandenberger Nato Asi series v 349

22 T Kibble and E Volovik in .ed Bunkov, Godfin Nato science series vol 6 2001

23 M Kastner Topological phase transitions arxiv

24 E Witten hep-th/9710065

25 Topological phasees papers LCastti,M Pettini,EGD Cohen cond -mat 0104267, cond-mat 0303200,cond-mat 0205483, A Okounkov, N Reshetikin,C Vafa hep-th 0309208 cond-matt 0205483, hep-th 0309208,hep-th0312022 , hep-th 0405146 ,hep-th 0004204;gr-qc 0403079

26 Gravity gauge theory papers  Nikolas Batakis gr-qc 9711054, T Ackermann,J.Tolksdorf hep-th 9503180 ,

27 J Samuel gr-qc 0005095 ,,hep-th 0206213, hep-ph 0208129 ,A.Magnon Ashtekar variables and unificationof gravitation and electromagnetism classical and quantum gravity v.9supplDec1992 s169-181.

28.M Carmelli WSP2001 Classical fields  (SU(2) gauge theory of gravity; M Carmelli , E Leibowitz ,N Nissani 1990 Gravitation SL(2,C) gauge theory ;G Sardanasshvily and Ozakharov, Gauge theory of gravitation

29 John Wheeler Geometrodynamics 1962;John Baez Gauge fields, knot theory and gravity WSP 1992; Reviews by C Rovelli , L Smolin, J Pullin, J Lewandowski on loop quantum gravity

30 Ajay Patwardhan quant-ph0305150; quant-ph0211041 ;  
      quant-ph0211039; gr-qc0310136

\end{document}